\documentclass{mn2e}

\usepackage[dvips]{graphicx}

\def\etal{{\rm et al. }}
\def\mpc{{h^{-1} \rm Mpc}}
\def\kpc{{h^{-1} \rm kpc}}
\def\kms{{\rm kms^{-1}}}

\title{Properties of galaxies in SDSS Quasar environments at $ z < 0.2$}

\author[G. V. Coldwell and D. G. Lambas]
       {Georgina V. Coldwell\thanks{On a fellowship from CONICET, 
        Argentina} and
        Diego G. Lambas\thanks{CONICET, Argentina}\\
        Grupo de Investigaciones en Astronom\'{\i}a Te\'orica y Experimental
        (IATE), Observatorio Astron\'omico,\\
        Universidad Nacional de C\'ordoba, Laprida 854, 5000, C\'ordoba,
Argentina.\\
        e-mail: georgina@oac.uncor.edu, dgl@oac.uncor.edu}

\date{\today}

\begin{document}
\pagerange{\pageref{firstpage}--\pageref{lastpage}}

\maketitle

\label{firstpage}

\begin{abstract}

We analyse the environment of low redshift, $z < 0.2$, SDSS quasars  
using the spectral and photometric information of galaxies from the Sloan 
Digital Sky Survey Third Data Release (SDSS-DR3). 
We compare quasar neighbourhoods with field and high density environments 
through an analysis on samples of typical galaxies and groups.

We compute the surrounding surface number density of galaxies finding that 
quasar environments systematically avoid high density regions. Their mean 
environments correspond to galaxy density enhancements similar to those of 
typical galaxies. 

We have also explored several galaxy properties in these environments, such 
as spectral types, specific star formation rates, concentration indexes, 
colours and active nuclei activity. 
We find a higher relative fraction of blue galaxies in quasar environments 
compared to groups and typical galaxy neighbourhoods. Consistent with this 
picture, the distribution of the concentration index of these galaxies also 
indicate a larger fraction of late-type objects. By analysing the available 
information of galaxy spectra we have also studied the distribution of the 
star formation rates of these neighbour galaxies finding that quasar  
environments are populated by objects with an enhanced star formation activity.
An analysis of the relative flux ratios of $[OIII]\lambda5700/H\beta$ and 
$[NII]\lambda6583/H\alpha$ of emission-line galaxies shows no 
an excess of nuclei activity in quasar neighborhood with respect to the 
environment of a typical galaxy. 

We conclude that low redshift quasar neighbourhoods ($r_p < 1\mpc$, $\Delta V < 500\kms$) 
are populated by bluer and more intense star forming galaxies of disk-type 
morphology than galaxies in groups and in the field. 
Although star formation activity is thought to be significantly triggered by 
interactions, we find that quasar fueling may not require the presence of a 
close companion galaxy ($r_p < 100\kpc$, $\Delta V< 350\kms$).

As a test of the unified AGN model, we have performed a similar analysis to
the neighbours of a sample of active galaxies. The results indicate that these neighbourhoods
are comparable to those of quasars giving further support to this unified scenario.

\end{abstract}

\begin{keywords}
quasars: active galaxies : statistics-- distribution --
galaxies: groups: general --
\end{keywords}

\section{Introduction}

Important clues on galaxy formation and evolution may be
obtained by characterizing statistically the properties of their
neighbourhoods in the local Universe.
It is well known that several properties of galaxies depend on the environment 
where they formed and evolved where a 
variety of processes such as star formation, tidal stripping, merging, etc. 
can determine the nature of galaxies. 
It should also be considered the possibility of AGN feedback processes
which could induce significant changes in the evolution of their
companion galaxies (see for instance Croton et al. 2005).

Previous works aimed to characterize the quasar neighbourhood indicate that high 
redshift quasars can be used as signspots to search for rich density regions 
(Djorgovski 1999, Hall \& Green 1998, Fukugita 2003).
However,
at lower redshifts (z $\le $ 0.3) different studies indicate that 
quasars could reside in environments similar to those of normal galaxies (Smith, 
Boyle \& Maddox 1995, Sorrentino \etal 2006) or they could be located in groups (Fisher et al. 1996) or 
clusters of galaxies (Mclure \& Dunlop 2001).
Coldwell et al. 2002, analysing both the projected cross-correlation function and colours
of galaxies around a sample of quasars and AGNs found that their typical galaxy density 
environment corresponds to groups of galaxies. Moreover, 
Coldwell \& Lambas (2003, hereafter PaperI), using the spectral information of 
the 2dF Galaxy Redshift Survey (2dFGRS), found that the galaxies within $r_p < 1 \mpc$ from 
quasars have a stronger star formation activity. 
These results are also supported by findings of Soechting \etal (2002, 2004) who found that low
redshift quasars follow the large-scale structure traced by galaxy clusters but they
 are not placed in the central area of galaxy clusters. More likely, they are in the cluster
periphery or between two, possibly merging, galaxy clusters.

The large recent spectroscopic surveys, the Sloan Digital Sky Survey and the 2dF Galaxy 
Redshift Survey allow us investigate a wide range of galaxy properties such as morphology, 
spectral types, colours and, also, the galaxy density using volume-limited samples. 
In this paper we extend the work of PaperI by analysing the environment of Sloan quasars 
using the spectroscopic and photometric information available for the Sloan Digital Sky 
Survey Third Data Release (SDSS). 
We divided the sample in two range of redshift $0.03 < z < 0.1$ and $0.1 < z < 0.2$ 
in order to anaelyse different effects in these two redshift ranges populated by galaxies of
different luminosities.

The layout of the paper is as follows.
In section 2 we describe the data, section 3 shows the analysis of the local density estimates,
section 4 and 5 describes the statistical
analysis performed with the photometric and spectroscopic data respectively, and in section 
6 we provide a brief discussion of the main results.

\section{Data}

The Sloan Digital Sky Survey (SDSS) in 5 optical bands will 
map one-quarter of the entire sky and perform a redshift survey of galaxies, quasars 
and stars. The third data release, DR3, from SDSS provides a database of $374767$ 
galaxies and $51027$ quasars with measured spectra. The five filters \textit{u, g, r, i,} 
and \textit{z} cover the entire wavelength range of the CCD response (Fukugita \etal 1996).
The main galaxy sample is essentially a magnitude limited spectroscopic sample (with a 
Petrosian magnitude) \textit{$r_{lim}$}$ < 17.77$, most of galaxies span a 
redshift range $0 < z < 0.25$ with a median redshift of 0.1 (Strauss \etal 2002).

The quasar sample is defined by quasars which have at least one emission line with a full
width at half maximum larger than $1000 \kms$, luminosities brighter than $M_i = -23$,
PSF magnitude $i < 19.1$ and highly reliable redshifts (Schneider \etal 2003).

The galaxy groups from SDSS were identified by Merch\'an \& Zandivarez (2005) by using
the friends-of-friends algorithm developed by Huchra \& Geller (1982) which was improved by
implementing a procedure to avoid the artificial merging of small systems in high density
regions and applying an iterative method to recompute the group centres position. The group
sample has a median velocity dispersion of $230 \kms$ and we restrict our target to
those groups with a minimum number of 8 members to obtain higher density environment.

As a way to reject the hypothesis that quasars could reside in environments such as that
corresponding to galaxy groups
we use three different target samples, taken from SDSS, in order to compare quasar 
environment with
those corresponding to typical galaxies and galaxy groups and
we make appropriate comparisons of galaxy characteristics 
of quasar neighbourhoods with respect to those in a low density environments
corresponding to typical galaxies and denser environments as galaxy groups.

We analyzed galaxy properties in the neighbourhood of different target samples
in two ranges of redshifts in order to explore for a luminosity dependence:

\begin{description}
\item [$0.02 < z < 0.1$, hereafter $Z1$:] 418 SDSS Quasars, 1147 SDSS Galaxies and 
                                          779 SDSS Galaxy Groups .

\item [$0.1 < z < 0.2$, hereafter $Z2$:] 1652 SDSS Quasars, 1153 SDSS Galaxies, 102
                                         Galaxy Groups. 
\end{description}

The targets were selected to match the observed quasar redshift distribution,
showed in Fig. 1,  
in order to have unbiased and directly comparable results and the samples have the
largest number of objects (within the redshift restriction) which suffices to
provide reliable statistical results. By doing so, we assume that there are not  
redshift-dependent systematics such as those associated to low luminosity galaxies. 
A Kolmogorov-Smirnov test yields that the redshift distributions are very similar
with a 90 \% level of confidence.

Tracer galaxies consist of all objects within projected distance $r_p < 3 \mpc$ and 
with radial velocity difference $\Delta V < 500\kms$ relative to the targets, so that
both targets and tracers have a similar redshift distribution.   

\begin{figure}
\includegraphics[width=72mm,height=75mm]{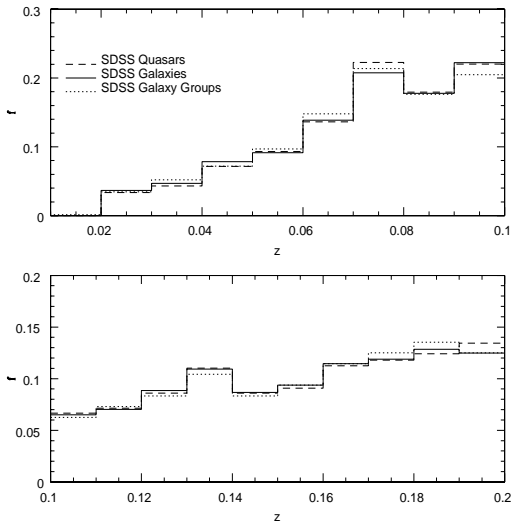}
\caption{Redshift distributions of targets in the redshift intervals $Z1$ and $Z2$.  Dashed 
lines correspond to quasars, solid lines to typical galaxies, and dotted lines to groups 
of galaxies.}
\label{fig1}
\end{figure}
 
\section{Local Density estimates}

A useful characterization of the local galaxy density can be obtained by measuring the 
distance 
to the $N^{th}$ nearest neighbour and estimating the density within that distance. 
The advantage of this method is to use a systematically
larger scale in lower-density regions which improves sensitivity and precision at
low densities.
This is a two-dimensional estimate but we use the redshift information to lower the
projection effects. We choose a fixed velocity interval of $\Delta V=1000\kms$ to compute
the local density which correspond to galaxies within $\sim 3\sigma$ from the centre
of a galaxy cluster (Balogh \etal 2004) and this allows inclusion of galaxies in system
with large velocity dispersion.
When computing projected distances we have assumed a flat cosmology
($q_0=0.5$, $\Omega=0.3$, $\Omega_{\lambda}=0.7$) and a Hubble constant $H_0=100h{\rm{kms^{-1}Mpc^{-1}}}$.

We calculate surface densities, $\Sigma_1$ and $\Sigma_5$, corresponding to the projected 
distance $d_1$ and $d_5$ of the first and fifth neighbour brighter than $M_r < -20.5$ 
respectively. This absolute magnitude limit assures completeness within the redshift range 
explored, $ z < 0.2 $
 
\begin{equation}
\Sigma_n=\frac{n}{\pi d_{N}^2} 
\end{equation}

\begin{figure}
\includegraphics[width=90mm,height=80mm]{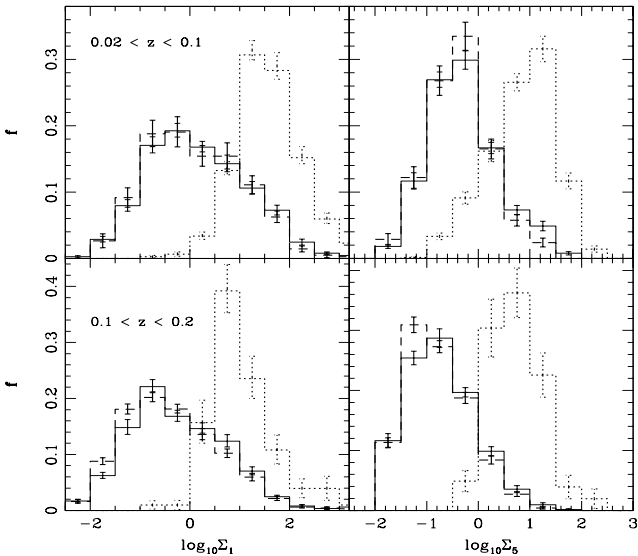}
\caption{Distributions of $log_{10}$ $\Sigma _1$ (left panels) and $log_{10}$ $\Sigma _5$
(right panels) for galaxies in the neighbourhoods of quasars (dashed lines), 
typical galaxies (solid), 
and groups of galaxies (dotted), for two ranges of redshift. 
}
\label{fig2}
\end{figure}
 
In Fig. 2 we show the relative distribution of $\Sigma_1$ and $\Sigma_5$ where it can be 
appreciated that the local
density of quasar environment is pretty similar to that corresponding to typical
galaxies. 
The $\Sigma_1$ density parameter corresponds to a closest neighbor distance estimate 
(within the luminosity restrictions) so that the results of Fig. 2 indicate that
galaxy interactions are not likely to be directly associated to quasar
phenomena.
The errorbar in this figure and in all figures were calculated by using 
boostrap error resampling (Barrow \etal 1984).
The results shown in Fig. 2 strongly suggest that local quasars avoid systematically 
high and moderately high density regions such as groups of galaxies.

\section{Photometric Properties}

Different magnitude measurements are provided in SDSS, asinh magnitude,
Petrosian magnitude, fiber magnitude, etc. where galaxies bright enough to be included
in the spectroscopic sample, $r<17.7 $, have relatively high signal-to-noise ratio.
Since Petrosian magnitudes are model independent and yield a large 
fraction of the total flux, roughly constant with redshift, they provide an 
adequate magnitude measurement. 
We have used the modified form of the Petrosian (1976) system 
which measures galaxy fluxes within a circular aperture whose radius is defined by
the shape of the azimuthally averaged light profile in order to measure a constant
fraction of the total light, independent of the position and distance of the objects.

\subsection{SDSS Colours}

Galaxy colours can be used as estimators of the galaxy evolution. 
In clusters, the large fraction of red galaxies
indicate an old population of galaxies with a low star formation rate. Galaxies
in poor groups or in the field are bluer and with stronger star formation rate.

\begin{figure}
\includegraphics[width=90mm,height=80mm]{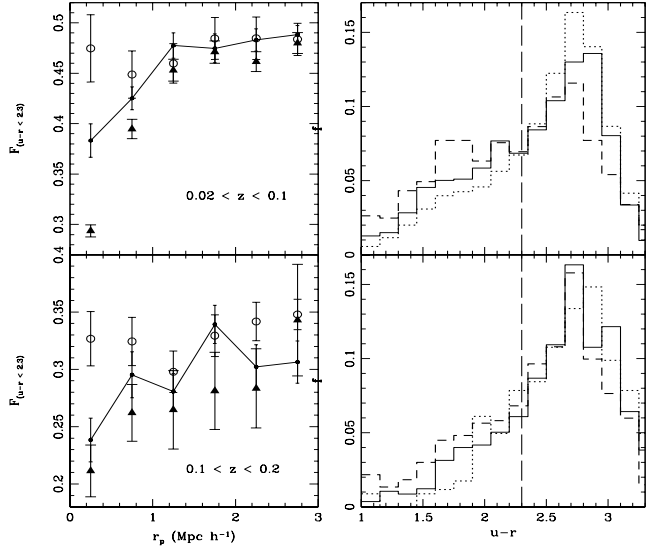}
\caption{Left panels: Fraction of blue galaxies, $u-r < 2.3$, relative to the 
total number in each bin of projected distance $r_p$ from the targets:
quasar neighbourhoods (empty circles), typical galaxies (solid lines), 
and groups of galaxies (filled triangles). Right panels: Distributions of $u-r$ 
colours of neighbour galaxies, within $r_p < 0.5 \mpc$, to Sloan quasars 
(dashed lines), groups of galaxies (dotted lines) and typical galaxies (solid line) in 
the range $0.02 < z < 0.1$, $b)$ $0.1 < z < 0.2$.
}
\label{fig3}
\end{figure}
 
We calculate the corrected colours $u-r$ of galaxies in the neighbourhood of the 
targets by using K and extinction corrected absolute magnitude (Blanton 2003).
We analyse the relative distribution of $u-r$ for galaxies around the
targets where we observe an excess of blue galaxies in the environments
of quasars. To determine the significance of this excess we calculate the relative
fraction of galaxies bluer than a given threshold value, we selected the fraction
of galaxies with  $u-r < 2.3$ which correspond approximately to the mean $u-r$ for
the spectroscopic survey at $z < 0.2$. The results
are shown in Fig. 3. 

We can see in this figure that the fraction of blue neighbours of a
typical galaxy tends to decrease at small separations. This effect that
can be interpreted as the usual morphology-density relation and, as
expected, this decrease is even stronger as one approach group centres.
However, contrary to this behaviour, the blue fraction does not decline at close
separations and there is a significant relative excess of blue galaxies
in quasar neighbourhoods compared to typical galaxies. This result is
perhaps an indication of feedback process operated by quasars which
could delay star formation at early stages allowing for more recent
activity from the remaining gas.

\subsection{Morphology}
In SDSS are also available the apparent radii containing $50 \% $ and $90\%$ 
of the Petrosian flux for each band.
The ratio of these fluxes, the concentration index $C$,
is correlated with morphology. Since galaxies with a de-Vaucouleurs profile have 
a value of $C \sim 3.3$ and disk galaxies have a concentration 
index $C \sim 2.4$ this parameter can be used as a simple morphological 
classifier.   
In Fig. 4 we can see the distribution of the concentration 
index parameter and the fraction of late type galaxies defined with $C < 2.5$
as a function of $r_p$. 
The larger relative fraction of disk-type galaxies in quasar neighbourhoods can be clearly 
appreciated corresponding to $\sim 20 \%$ excess of disk-type objetcs.

\begin{figure}
\includegraphics[width=90mm,height=80mm]{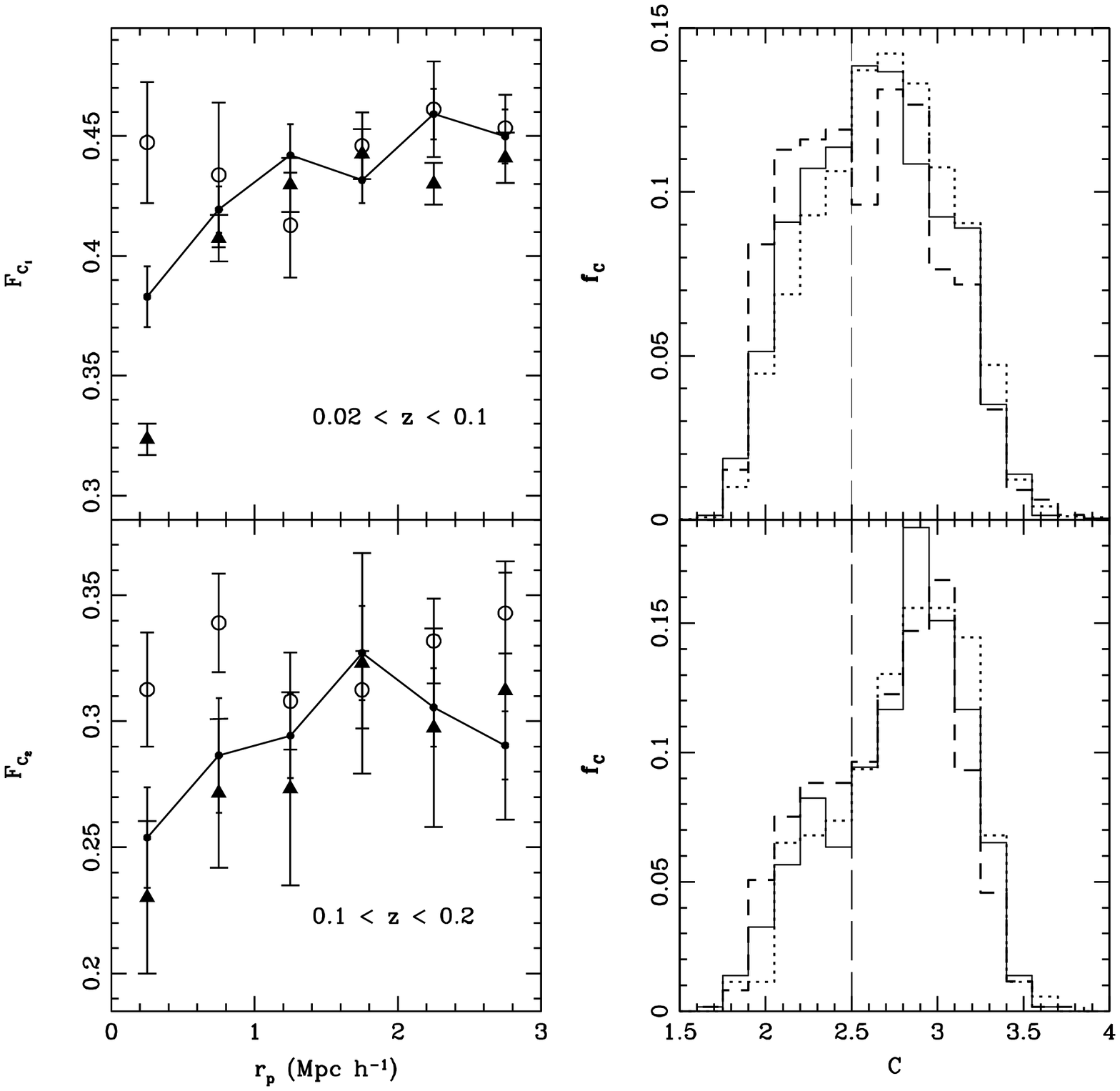}
\caption{Left panels: Fraction of disk-type galaxies, $C < 2.5$, relative to the 
total number in each bin of projected distance $r_p$ from the targets:
quasar neighbourhoods (empty circles), typical galaxies (solid lines), 
and groups of galaxies (filled triangles). Right panels: Distributions of 
concentration index C of neighbour galaxies, within $r_p < 0.5 \mpc$, to Sloan quasars 
(dashed lines), groups of galaxies (dotted lines) and typical galaxies (solid line) in 
the ranges Z1 and Z2.
}
\label{fig4}
\end{figure}

\section{Spectral properties}
\subsection{Spectral Type Classification}

The regulation of star formation in galaxies can be strongly influenced by close 
companions. Besides the effects of environment, galaxies in pairs have enhanced 
star formation over a control sample with similar characteristics which is
stronger for galaxy pairs in field that in groups of galaxies.
 Lambas et al. (2003) and Alonso et al (2004) found clear evidence that star 
formation is enhanced over $40\%$ in close interactions. This star formation 
induction increases for small relative velocity and projected separation $r_p$.

On the other hand, feeding  a massive black hole at the centre of galaxies
may require particular conditions which could be strongly influenced by 
environment. Although at large redshifts quasars and radio galaxies are 
frequently associated to clusters, at lower redshifts, a dense environment
may be hostile to the presence of massive black holes (Coldwell, Martinez 
\& Lambas 2002, Coldwell \& Lambas 2003).

\begin{figure}
\includegraphics[width=90mm,height=80mm]{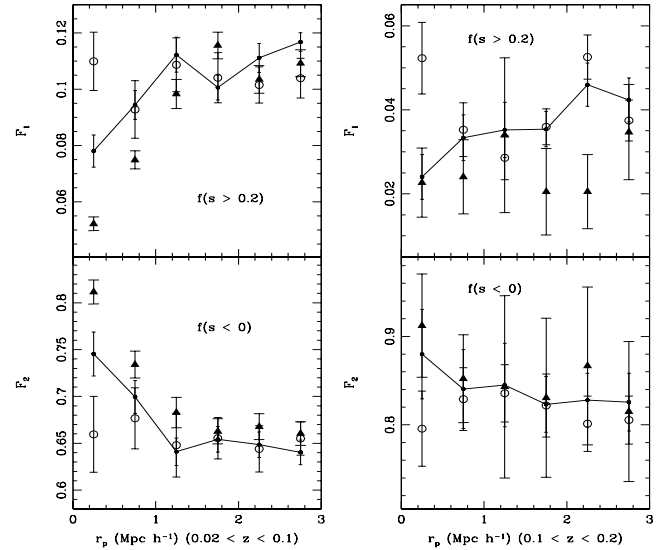}
\caption{Fraction of emission-line galaxies ($s > 0.2$), $F_1$,
and fraction of early galaxies ($s < 0$), $F_2$, relative to the total number in each 
bin of projected distance $r_p$ from the targets in the 
neighbourhoods of quasars (empty circles), typical galaxies (solid line)
and groups of galaxies (filled triangles) in the ranges $Z1$ (left panels) and $Z2$ 
(right panels).
}
\label{fig5}
\end{figure}

In PaperI we explored the nature of galaxies
in the vicinity of the different target samples using the 2dFGRS spectral type index $\eta$. 
This parameter approximately delineates the transition between early and late 
morphological types for $\eta \simeq -1.4$ . When considering $\eta > 3.5$ we are 
dealing with galaxies particularly active, such as starbursts with recent episodes of 
star-formation or AGNs.

SDSS galaxies are also classified by a principal component analysis, PCA 
(Connolly \& Szalay 1999), where five eigencoefficients, ecoeff, are extracted.
Similarly to the 2dFGRS spectral type classification SDSS provides a spectral
type parameter s with the first two eigencoefficients like $s=arctan(-ecoeff2/ecoeff1)$
which ranges from about -0.35 to 0.5 for early to late galaxies.
Taking in to account this and the corresponding galaxies
for SDSS and 2dFGRS, we consider galaxies with $s < 0$ as being similar spectra 
than their 2dF counterparts with ($\eta < -1.4$). Also, SDSS 
galaxies with $s > 0.2$ correspond to the 2dF galaxies with $\eta > 1.1$
The fraction, $F_1$, galaxies with strong star formation activity ($s > 0.2$), 
and the fraction $F_2$ of early spectral type galaxies ($s < 0$), are shown for the
two ranges of redshifts $Z1$ and $Z2$ in Fig. 5.
The effect is similar to that of 2dF galaxies in Fig. 1 of PaperI where it is clear that
galaxies around quasars differ from that in galaxy and group environments.

\subsection{Star formation vs. AGN activity}

Several important properties of galaxies have been derived for subsamples of SDSS:
 stellar masses, indicators of recent major starbursts, current total and specific 
star-formation rates (Brinchmann \etal 2004) and emission-line fluxes (Tremonti \etal 2004) 
both for the regions with spectroscopy and for the galaxies as a whole; gas-phase 
metallicities; AGN classifications (Kauffmann \etal 2003) 
based on the standard emission line ratio diagnostic diagrams, etc.

The relation between spectral lines, $[OIII]\lambda 5007$, $H\beta$, 
$[NII]\lambda 6583$ and $H\alpha$ luminosities, can be used to analyse possible 
dependence of the relative numbers of AGNs and star forming galaxies in quasar 
neighborhoods compared to galaxy and group environments.
We constructed the Baldwin, Phillips \& Terlevich (BPT, 1981) line-ratio diagram    
and we used the Kauffmann \etal (2003) criteria to differentiate AGNs galaxies
from other emission-line objects dominated by star formation. An AGN is defined if 

\begin{equation}
log([OIII]/H\beta) > 0.61/(log([NII/H\alpha])-0.05)+1.3
\end{equation}

Figure 6 shows the BPT diagram for redshift range Z1 where it can be appreciated a lack of
differences in the distribution of AGN neighbours for quasars, galaxies and AGNs.
Similar results were found for the redshift range Z2. 

To quantify this result we calculated the percentage of AGN neighbours within $1\mpc$ 
from the three target samples obtaining that 37.9 \% , 35.2\% and 34.5 \% 
of galaxies in the vicinity of groups, galaxies and quasars, respectively,
are AGNs indicating that the AGN activity is not strongly affected by quasars.

\begin{figure}
\includegraphics[width=80mm,height=90mm]{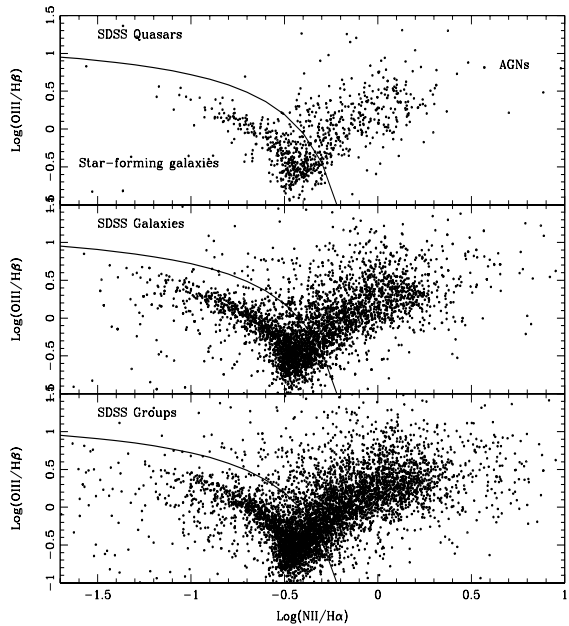}
\caption{The distribution of the neighbour ($r_p < 1\mpc$) galaxies in the BPT line-ratio 
diagram for 
the three target samples at Z1. The solid line shows the division of the samples.   
}
\label{fig6}
\end{figure}

Taking into account the previous results which indicate that quasar
environments are populated by blue, disk-type, star forming galaxies, we
have computed the fraction of AGNs (in different environments) 
which have these characteristics ($g-r < 0.7$ and  
$C < 2.5$). 
The corresponding fractions of AGNs obtained are 6.7 \% , 5.3\% and 6.4 \% 
in the vicinity of groups, galaxies and quasars, respectively indicating that AGN 
contribution to blue star forming galaxies is not significant in any environment.

\subsection{Star Formation Rates}

The methods for deriving the star formation rates are based on models where
the contribution of the nebular emission by HII regions and diffuse
ionized gas combined and described in terms of effective metallicity,
ionization parameter, dust attenuation at 5500 $\r{A}$, and dust to metal 
ratio (Bruzual \& Charlot 1993, Charlot \etal 2002). 
Taking into account these issues, Brinchmann \etal (2004) provide accurate total
star formation rates estimates free from aperture bias.
Moreover, Kauffmann \etal (2003) developed a method to constrain star 
formation histories, dust attenuation and stellar masses of galaxies based
on two stellar absorption lines indices, the 4000 $\r{A}$ break strength
and the Balmer absorption line index $H\alpha_A$.

Our main interest here is to analyse the logarithmic specific star formation 
rate $log SFR/M^*$ [$log$ yr$^{-1}$], where M* is the estimated mass in star, 
for galaxies in the different target environments. 
In the right panels of Fig. 7 it is shown the distribution of  
 $log SFR/M^*$ from which it can be seen
that quasar environments have a population of 
galaxies with a higher star formation rate than field galaxies
(galaxy groups show a low rate of star formation as expected from the systematic 
decline of SFR with local density).

\begin{figure}
\includegraphics[width=90mm,height=80mm]{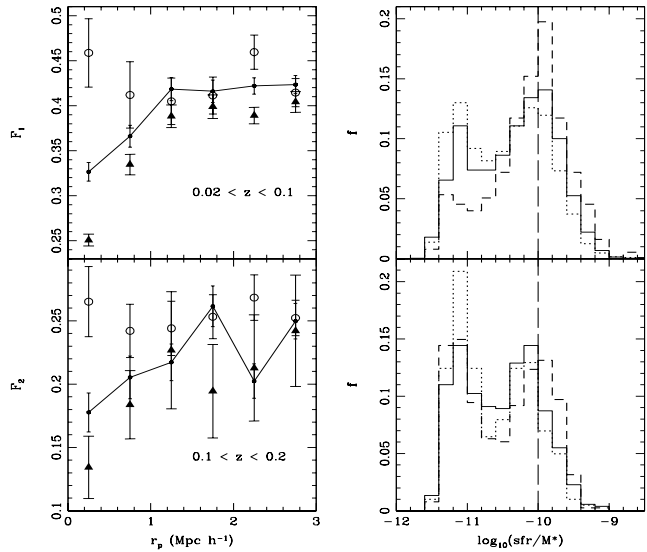}
\caption{Left Panels: Fraction of galaxies with strong specific star formation rate, 
$log SFR/M^* > -10.0$, 
relative to the total number in each bin
of projected distance $r_p$ from the targets in the 
neighbourhoods of quasars (empty circles), typical galaxies (solid line)
and groups of galaxies (filled triangles)
Right Panels: Distribution of star formation rate per mass unit, $log SFR/M^*$,
within $r_p < 0.5\mpc$ from the targets in the 
neighbourhoods of quasars (dashed line), typical galaxies (solid line)
and groups of galaxies (dotted line) .
}
\label{fig7}
\end{figure}

In order to quantify the excess of star formation activity, we calculate the fraction of 
galaxies
with $log(SFR/M^*) > -10.0$ and the results are shown in Fig. 7 (left panels).
Several theories have proposed that 
galaxy-galaxy interactions fuel the AGN activity by driving gas into the cores
of galaxies and thus onto the black holes.
We have directly explored whether quasars have companions more frequently
than the control samples. Taking into account the observed thresholds in projected 
separation $r_p$ and relative velocity $\Delta V$ for galaxy interactions to 
effectively induce star formation we calculated 
the fraction of targets with close companions ($r_p <100\kpc$ and
$\Delta V< 350 \kms$). We find a low fraction ($\le 15 \%$) of
close companions associated to quasars, similar to that observed for 
galaxies in general. This result indicates that quasar fueling may not require
the presence of close companion galaxies (consistent with the distribution
of $\Sigma_1$ shown in Fig. 2). Moreover we have calculated the distributions of
star formation excluding these close neighbours (those within $r_p <100\kpc$ and
$\Delta V< 350 \kms$). The results are very similar 
to those of Fig. 7, indicating the lack of relevance of close interactions in
driving these relations.

\subsection{Comparison of Quasar and AGN environments.}

The unification hypothesis for active galactic nuclei and quasars (Antonucci 1993)
is a model where all active galaxy classes are fundamentally the same phenomenon seen at 
different orientations which are not due to intrinsic differences 
As a final analysis we have explored the characteristics (colours, specific star formation 
rates and concentration indexes) of galaxies in AGN environments 
(ie. taking AGNs as targets) 
in a similar fashion as in previous sections serving as a test of the unified AGN model. 
The AGN target samples were taken from Kauffmann \etal (2003) restricted to have a similar
redshift distribution than the quasar, group, and typical galaxy samples analysed
previously. The results are shown in Fig. 8 where it can be 
appreciated the similarity of galaxy properties surrounding quasars and AGNs. 
The fact that the environments of quasars and AGNs are comparable provide 
further support to the unified model.  
We notice, however, that AGN environments more closely resemble that of quasars in the 
high redshift range, a fact that may be related to the higher luminosity of
 AGNs here.

\begin{figure}
\includegraphics[width=80mm,height=90mm]{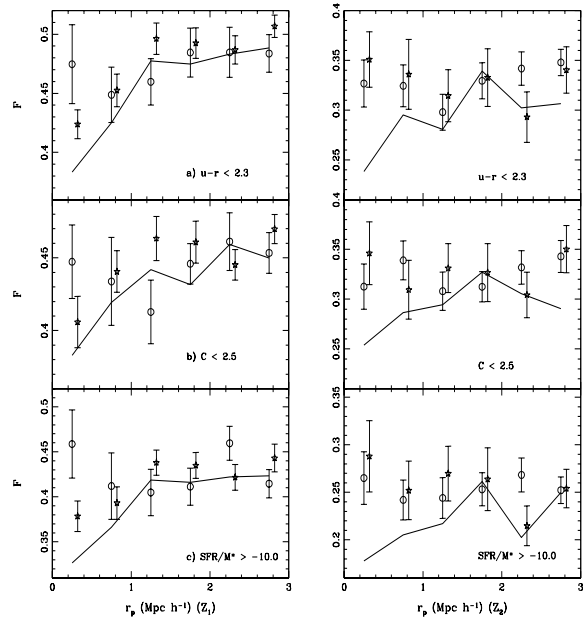}
\caption{Comparison of quasar and AGN environments. Left panels correspond to the redshift 
interval $Z1$ and  right panels to $Z2$. a), b) and c) correspond to the fractions of
blue colour index galaxies ($u-r < 2.3$), disk-types ($C< 2.5$), and strong specific star 
formation rate objects ($log SFR/M^* > -10$) respectively.
The empty circles correspond to quasar environments and the stars to AGN environments. The
solid lines show the results of typical galaxy neighbourhoods. 
}
\label{fig8}
\end{figure}

\section{Summary}

We have performed a statistical analysis of local quasar environments using SDSS data, 
our main conclusions are:

$\bullet$ Nearby quasars systematically avoid high density regions. Local surface density
estimates show that SDSS quasars reside in similar density enhancements to typical  
galaxies, significantly less dense than cluster environments.

$\bullet$ Star formation activity in the 
surrounding of quasars is higher than in the neighbourhood of typical galaxies.
This important property may provide a link between star formation and the onset of quasar
activity.

$\bullet$ Quasar environment is populated by galaxies systematically bluer,  
and with a disk-type morphology.

$\bullet$ In spite of the fact that galaxy interactions efficiently trigger star formation, 
we find no statistical evidence that the presence of close companions is associated to 
quasars.

$\bullet$The presence of quasars does not affect the fraction of AGNs in neighbour galaxies.

$\bullet$ The results mentioned are present in the two redshift ranges explored $z<0.1$ and 
$0.1<z<0.2$ although they are stronger at lower redshifts, indicating that the brightest 
galaxies are not likely to be the major contributors to the effects. 
Also, there might be a signal for evolution in the sense that the effects are stronger locally. 

$\bullet$ AGN and quasar environment are similar regarding the galaxy characteristics
explored and agree best in the highest redshift interval giving support to the unified
scenario of AGNs and quasars.

Although quasar hosts tend to be red and bulge-dominated galaxies (Kauffmann \etal 2003),
the indication of a relative excess of quasar neighbour
galaxies with a large gas fraction could suggest either an 
external origin for the accretion onto the central black hole, 
or that quasars may affect substantially
the evolution of star formation up to the Mpc scale.

\section{Acknowledgments}

We thank the anonymous Referee for helpful suggestions and comments.
This research was partially supported by grants from  CONICET,
Agencia C\'ordoba Ciencia and the  Secretar\'{\i}a de Ciencia y T\'ecnica
de la Universidad Nacional de C\'ordoba.

\label{lastpage}

\end{document}